\documentclass[letterpaper, 10 pt, conference]{ieeeconf}  

\IEEEoverridecommandlockouts                             

\usepackage{graphics} 
\usepackage{epsfig} 
\usepackage{mathptmx} 
\usepackage{times} 
\usepackage{amsmath} 
\usepackage{amssymb}  
\usepackage{graphicx}
\usepackage{subcaption}
\usepackage{multirow}
\usepackage{booktabs}
\usepackage{comment} 
\usepackage{soul, color} 
\usepackage{adjustbox}
\usepackage{url}

\title{\LARGE \bf Next-Generation Teleophthalmology: AI-enabled
Quality Assessment Aiding Remote
Smartphone-based Consultation
}

\author{
  D. Srikanth\thanks{\textsuperscript{1}Indian Institute of Technology Hyderabad, Telangana, India.}\textsuperscript{1},
  J. Gurung\thanks{\textsuperscript{2}L V Prasad Eye Institute, Hyderabad, Telangana, India.}\textsuperscript{2},
  N. S. Deepika\textsuperscript{1},
  V. Joshi\textsuperscript{2},
  L. Giri\textsuperscript{1},
  P. Vaddavalli\textsuperscript{2},
  S. Jana\textsuperscript{1}
}

\begin{document}

\maketitle

\thispagestyle{empty}
\pagestyle{empty}

\graphicspath{{}{images/}}

\begin{abstract}

Blindness and other eye diseases are a global health concern, particularly in low- and middle-income countries like India. In this regard, during the COVID-19 pandemic, teleophthalmology became a lifeline, and the Grabi attachment for smartphone-based eye imaging gained in use. However, user-captured images do not often possess complete clinical information for decision making, leading to delays. In this backdrop, we propose an AI-based quality assessment system with instant feedback mimicking clinicians' judgments and tested on patient-captured images. Dividing the complex problem hierarchically, here we tackle a nontrivial part, and demonstrate a proof of the concept.

\end{abstract}


\section{{Introduction}}

Early detection of sight-threatening disorders and prompt treatment help preserve visual function and potentially prevent blindness \cite{prevblindness}. Unfortunately, remote rural areas are often more severely affected. In overcoming access barriers and preventing delays (especially during the COVID-19 pandemic), smartphone-based teleconsultation has proven effective. In general, teleophthalmology not only increases the reach but also reduces treatment/other costs, conserves hospital resources and helps reduce the carbon footprint \cite{cfoot}. Now, can teleophthalmology provide outcomes similar to those traditionally obtained in hospitals?  

\begin{figure}[!t]
\begin{center}
\begin{tabular}{c}
\includegraphics[width=0.28\textwidth]{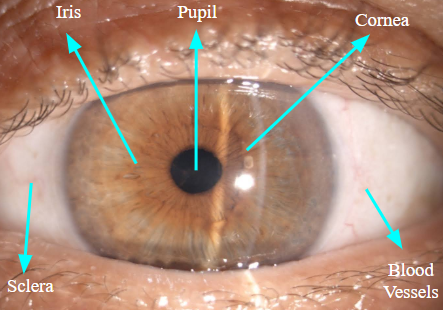}\\
(a)
\end{tabular}
\begin{tabular}{cc}
\includegraphics[width=0.2\textwidth]{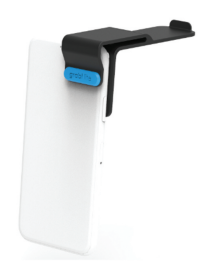}
        &
\includegraphics[width=0.14\textwidth]{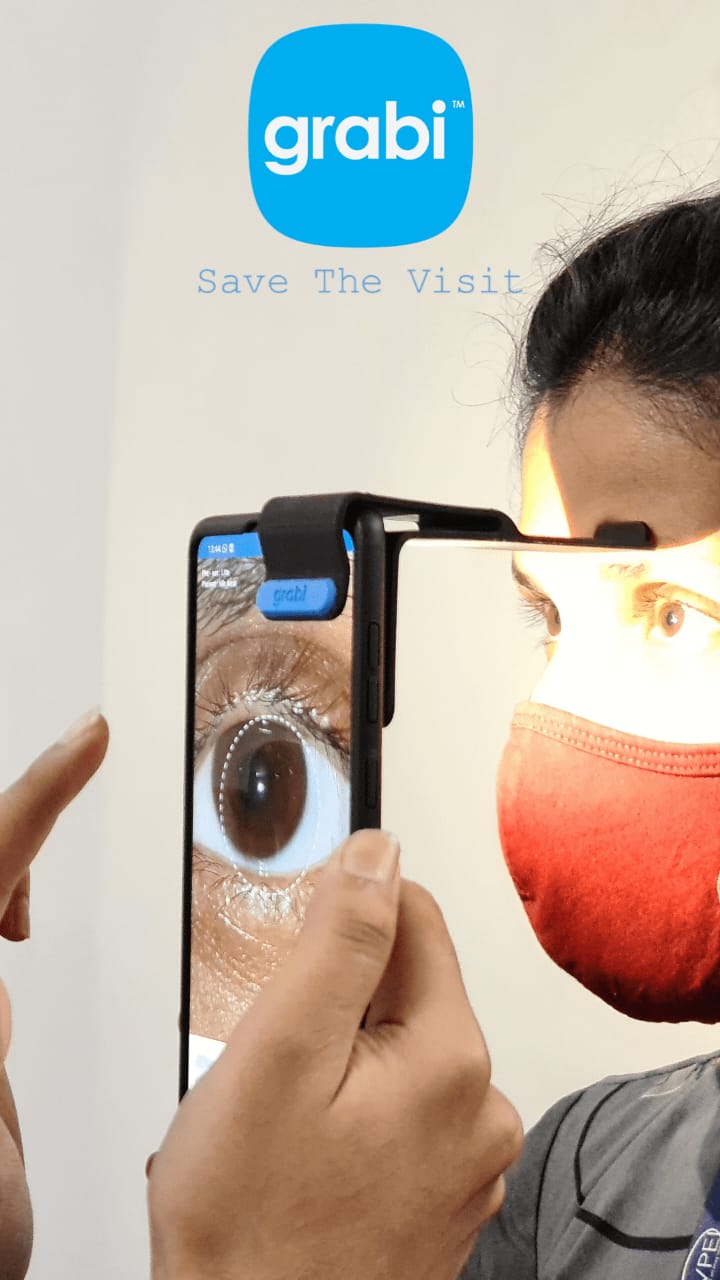}
        \\
        (b) & (c)
        \\       
\includegraphics[width=0.19\textwidth]{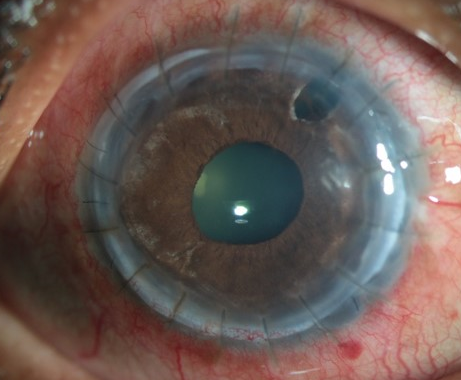}
        &
\includegraphics[width=0.1825\textwidth]{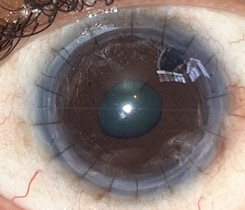}
        \\
        (d) & (e)
        \end{tabular}
\end{center}
    \caption{
    (a) Labelled slitlmap image of anterior segment of human eye, (b) Grabi universal  attachment fixed on dummy smartphone, (c) Grabi assisted eye image capture,  
    (d) and (e) Slitlamp and Grabi images, respectively, of the same eye.}
    \label{fig:Grabi}
\end{figure}

During a patient's physical  visits, an ophthalmologist usually requires high-quality slitlamp images of the eye captured by trained operators, such as Fig. \ref{fig:Grabi}(a), for diagnosis and treatment response monitoring. Thus, assuming an in-person prior visit and availability of reference slitlamp images, teleophthalmology requires mechanisms to ensure sufficient quality in smartphone-based self-captured eye images so that treatment response can reliably be monitored. As an aid, a custom-made app-based universal smartphone adapter, Grabi$^{\text{TM}}$, has recently been reported \cite{grabi}. As shown in Figs. \ref{fig:Grabi}(b) and \ref{fig:Grabi}(c), it facilitates capture of suitably magnified well-illuminated images of the anterior segment of the eye by maintaining appropriate direction and distance. As seen in Figs. \ref{fig:Grabi}(d) and \ref{fig:Grabi}(e), a Grabi image can be almost as informative as a slitlamp image. While patients were trained during hospital visits and found to gradually improve at image capturing, not all Grabi images turned out to be of satisfactory quality. Clinicians needed to vet those images and sometimes request re-capture, wasting time and effort. As a potential remedy, we propose to automate the process by assessing the quality of Grabi eye images using artificial intelligence (AI) algorithms so that the patient obtains an immediate feedback on the need to re-capture.


\section{Materials and Methods}
\label{sec:MM}

\subsection{Problem Statement and Approach}
\label{sec:close}

Clinicians assess quality of an anterior segment image of the eye based on various considerations. (1) Presence of eye: A satisfactory image must include an open eye. (2) Lighting: A satisfactory image must be adequately lit, and should not suffer from large variations in pixel intensity, low average pixel intensity, and visible reflections of the surrounding environment (possibly due to improper camera flash). (3) Resolution of cornea: The distance between the smartphone and the eye should be such that the eye in the image appears in sufficient size. (4) Completeness of cornea: The Cornea, the region of maximum clinical importance, should be fully represented in the captured image. (5) Focus: Key regions such as pupil periphery, iris periphery, iris striations, conjunctival blood vessels, and artificial structures such as sutures and grafts should appear distinct and sharp, even when certain eye details may be (inherently) blurry. In this paper, the quality being high is considered a positive event. A false positive, i.e., a low-quality image being incorrectly inferred as high-quality, triggers a re-take after costly clinician intervention, while a false negative directly triggers an unnecessary (but not costly) re-take by the patient.

\begin{figure}[t!]
    \centering         \includegraphics[width=\columnwidth]{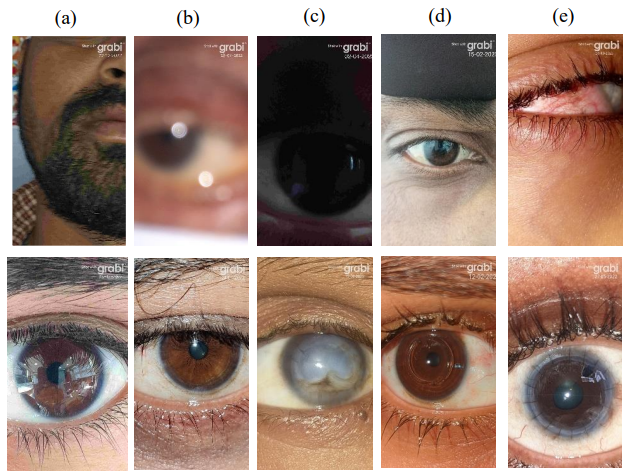}
    \caption{Ten representative Grabi images of the anterior segment with effect of specific factors illustrated columnwise: (a) presence of eye, (b) focus, (c) illumination, (d) magnification, (e) completeness of cornea; top panel  -- unsatisfactory images, bottom panel -- satisfactory images.}
    \label{fig:samp}
\end{figure}

\begin{figure}[!t]
    \centering
\includegraphics[width=0.35\textwidth]{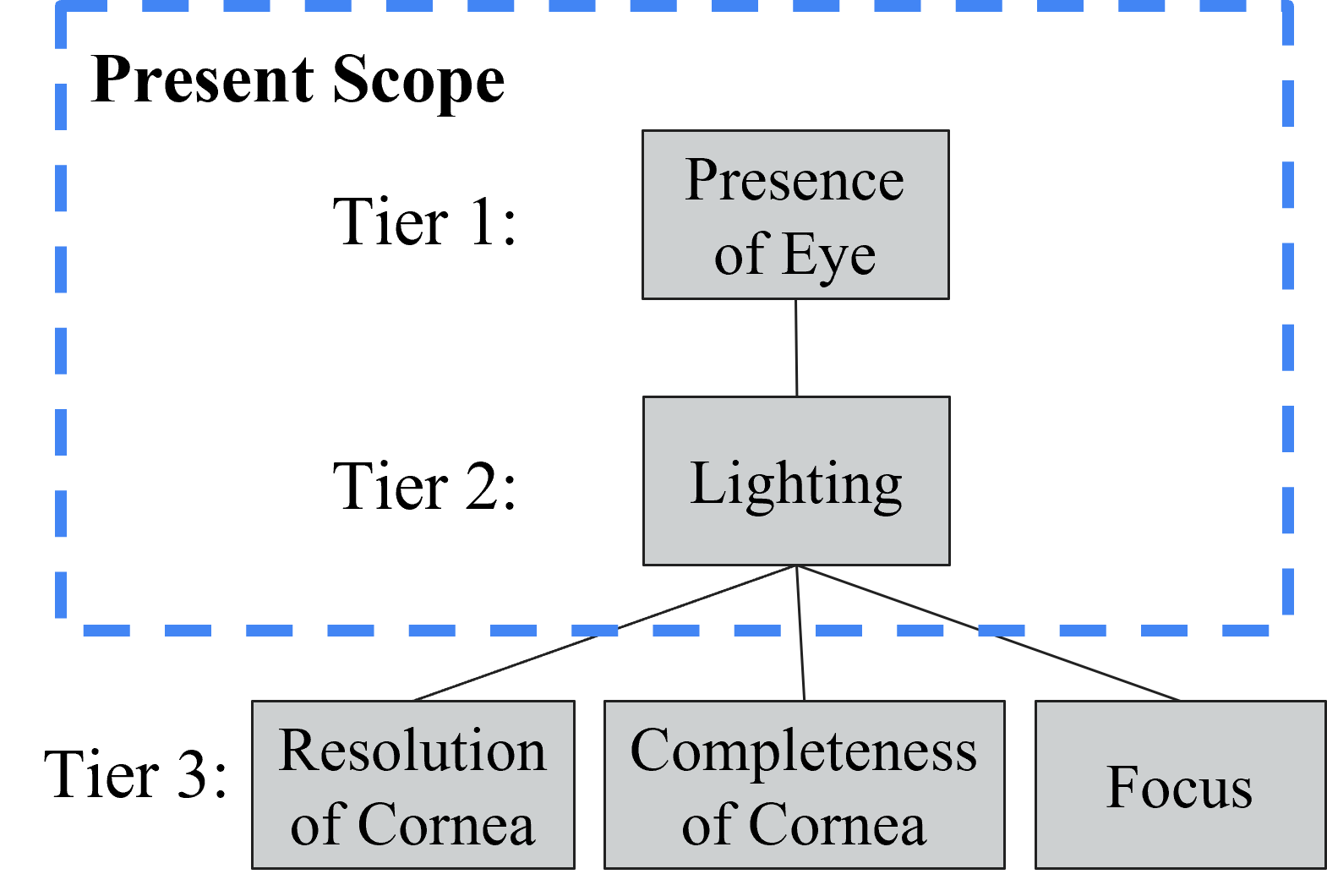}
\caption{Quality assessment of self-captured Grabi images: Hierarchical solution approach.}
\label{fig:layer}
\end{figure}

\begin{figure*}[t!]
    \centering        
    \includegraphics[width=0.85\textwidth]{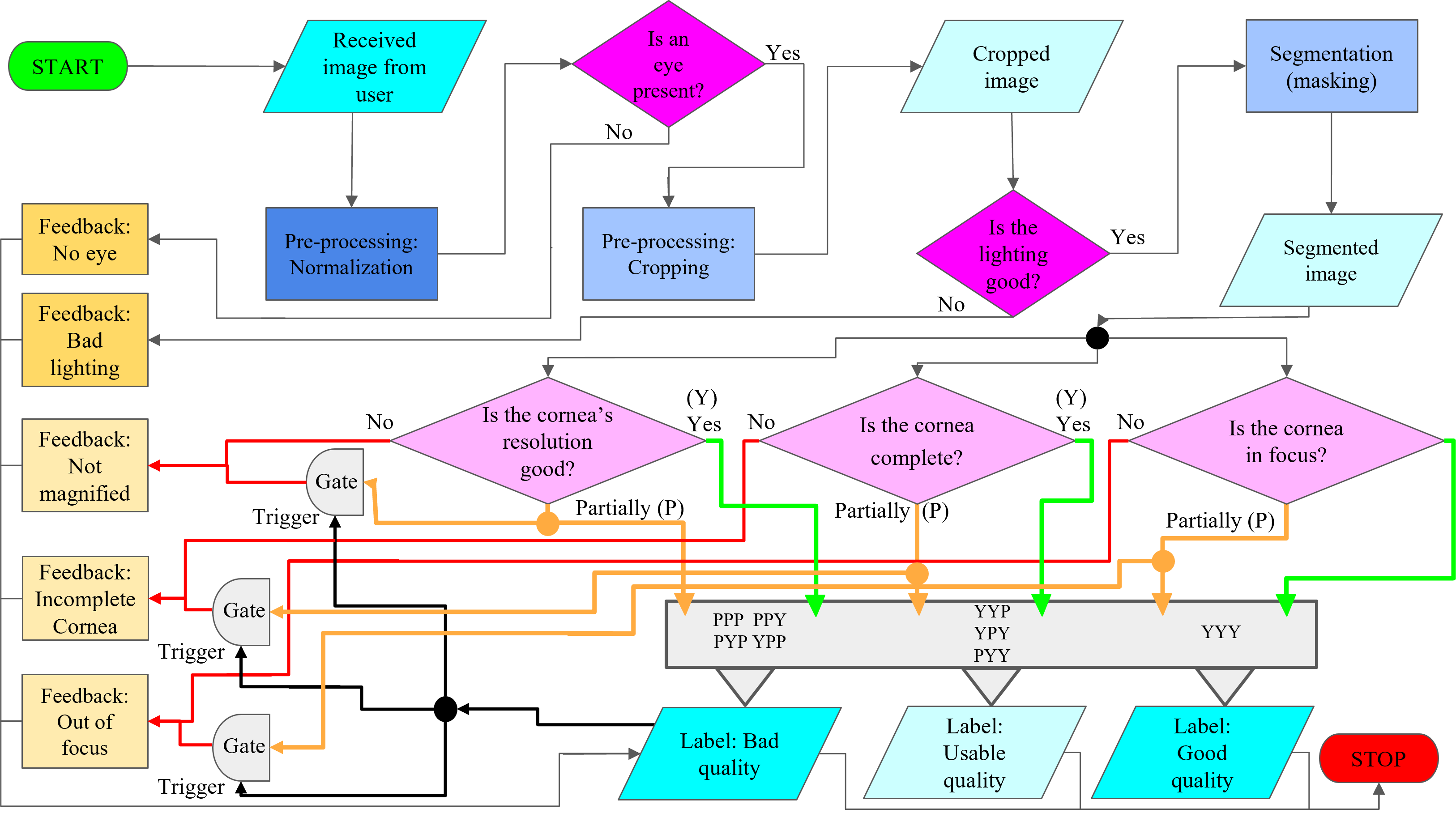}
    \caption{Flowchart of the envisaged solution for end-to-end image quality assessment.}
    \label{fig:fullflow}
\end{figure*}

At this point, refer to Fig. \ref{fig:samp} for
representative unsatisfactory (top row) and satisfactory (bottom row) images captured by Grabi-assisted smartphone camera corresponding to each of the above considerations (columnwise).
However, clinicians place certain priorities on those considerations, and do not treat those as equally important. Presence of an eye is the most important consideration in the sense that an image is unsatisfactory without it. If the eye is present, lighting takes the next priority so that an inadequately illuminated image is deemed unsatisfactory. Finally, an image with a well-lit eye present is considered usable if at least two out of the remaining three considerations are fully met, while the third consideration is met only per a relaxed standard.  


Now, our task boils down to mimicking clinicians' assessment of image quality using AI algorithms. To this end, we replicate the aforesaid hierarchy in clinical logic, and propose to verify presence of an eye in Tier 1, adequate illumination in Tier 2, and adequate resolution, completeness of cornea and satisfactory focus in parallel in Tier 3, as depicted in Fig. \ref{fig:layer}. Presently, we take up Tiers 1 and 2 only, deferring Tier 3 as future work.
A detailed flowchart is presented in Fig. \ref{fig:fullflow} (with parts related to the future Tier 3 dimmed).





\subsection{Data Collection and Data Preparation}
\label{sec:data}

Either the patient or an associate captures an image of the desired eye at home using the Grabi attachment and the Grabi app on a smartphone, and uploads it to the Grabi cloud. From there, we randomly picked images of varying quality to populate a dataset to be used in present experimentation. Here, we ignored metadata such as camera parameters/specifications, image resolution, aspect ratio, flash usage, and imaging environment. Each constituent image was labelled either satisfactory (positive) or unsatisfactory (negative) in relation to presence of eye, as well as separately to suitable lighting. The presence-of-eye (Tier 1) and suitable-lighting (Tier 2) detectors were developed separately, using 5200 (3700 with the eye present) and 4100 (3000 suitably illuminated) images, respectively. Each respective dataset was split
in the ratio 8:1:1 into training, validation and test subsets. For Tier 1, for example, we had 4160 training,  
520 validation  and 520
test images (2960, 370 and 370, respectively, with eye present). 
Each image was resized (or, zero-padded) to 224$\times$224, mean subtracted and normalized as depicted in Fig. \ref{fig:fullflow}. 

The aforementioned detectors were tested hierarchically, with the Tier 2 detector operating only on images classified as positive (eye present) by the Tier 1 detector. Analogously, a separate test set of 300 images was also labeled hierarchically: 100 without eye present, and of the 200 with the eye present, 100 each had suitable 
and unsuitable illumination.


\subsection{Proposed AI-based Tools and Solutions}
\label{sec:soln}

\subsubsection{General engineering principle} Both the presence-of-eye and the suitable-lighting detectors were realized using a particular residual CNN (convolutional neural network) architecture with 2 output nodes in each case \cite{resnet}, in view of its reported efficacy in classification tasks and relatively light weight. Specifically, as our base model of either detector, we used a ResNet-18 model pretrained on ImageNet benchmark object detection dataset, because of the representation of the home environment at which present eye images are expected to be captured. For each detector, suitable layers of the base model were trained using respective training data. More generally, each detector was individually developed and tested using the corresponding dataset, via the sequential process of training, validation and testing. Finally, those detectors were tested in the intended hierarchical configuration on another test set specially constructed for this purpose.

\subsubsection{Specific considerations and variants} 
We considered two variants of the presence-of-eye detector: Pertaining to the base ResNet-18 model, (i) only weights at the final layer are learnt keeping other weights at their pretrained values, and (ii) all weights/parameters are learnt. Two variants of the suitable-lighting detectors too were considered: Input data underwent (i) no transformation, and (ii) 
2-level Haar wavelet transform \cite{wavelet}. In either variant, all weights of the base ResNet-18 model were learnt afresh.

\begin{table*}[t!]
    \centering
    \caption{Independent testing results for 'presence of eye' and 'lighting' features}
    \begin{tabular}{cccccccc}
    \toprule 
        \textbf{Dataset} & \textbf{Parameters} & \textbf{Transforms} & \textbf{Accuracy} & \textbf{P4 Score} & \textbf{Custom Score} & \textbf{Learning} & \textbf{Momentum}\\ 
        & \textbf{Trained} && ($Mean \pm Std Dev$)&($Mean \pm Std Dev$)&($Mean \pm Std Dev$)& \textbf{Rate} &\\
    \midrule
        Eye Present & Last FC Layer & None & 96.532\% $\pm$ 0.6332\% & 0.957 $\pm$ 0.0082 & 0.951 $\pm$ 0.0144 & 0.0002 & 0.95\\
        & All Layers & None & \textbf{96.840\% $\pm$ 0.4656\%} & \textbf{0.961 $\pm$ 0.0061} & \textbf{0.957 $\pm$ 0.0071} & 0.0001 & 0.99\\
    \midrule
        Lighting & All Layers & None & \textbf{88.193\% $\pm$ 1.2096\%} & \textbf{0.849 $\pm$ 0.0180} & \textbf{0.862 $\pm$ 0.0391} & 0.0002 & 0.99\\
        & All Layers & Wavelet & 88.145\% $\pm$ 1.0492\% & 0.841 $\pm$ 0.0185 & 0.814 $\pm$ 0.0385 & 0.0002 & 0.99\\
    \bottomrule
    \end{tabular}
    \label{tab:pres}
\end{table*}

\subsubsection{Performance Metrics}
Further, while we make use of 
$Accuracy = {(TP + TN)}/{(TP + TN + FP + FN)}$ as a performance metric
($TP$, $TN$, $FP$ and $FN$, respectively, being the counts of true positive, true negative, false positive and false negative cases), it presently conveys limited meaning due to (i) possible class imbalance and (ii) the fact that a positive case being misclassified as a negative one (prompting unnecessary image retake by user) is less costly that the other way round (prompting significant wastage of clinician's as well as patient's time). In contrast, the $P4$ metric \cite{p4}, defined by
$$\frac{4}{P4} = {\frac{1}{Precision} + \frac{1}{Recall} + \frac{1}{Specificity} + \frac{1}{NPV}},$$
i.e., the harmonic mean of 
$Precision = {TP}/{(TP + FP)}$, 
$Recall = {TP}/{(TP + FN)}$, $Specificity = {TN}/{(TN + FP)}$ and negative predictive value ${NPV} = {TN}/{(TN + FN)}$, is insensitive to data imbalance and relative weightage of positive and negative classes. However, in order to cater specifically to issue (ii), in addition to (i), we coined a custom metric defined by the harmonic mean of $Precision$
and $Specificity$
$$\frac{2}{Custom} = \frac{1}{Precision} + \frac{1}{Specificity},$$
which de-emphasizes $FN$ (that simply allows image retake).

\subsubsection{Training, validation and testing} As mentioned earlier, each of the aforementioned detectors were trained, validated and tested on the corresponding dataset with a 8:1:1 split. Indeed, the above splitting was performed 5 times randomly, stratified by label. Each time, model parameters/weights were learned based on the training subset by minimizing cross-entropy loss using stochastic gradient descent. Further, after each epoch, training and validation losses were recorded with the aim of picking the parameters learnt at the epoch with the minimum validation loss, to prevent overfitting. 

For the parameters so chosen, we considered the detector validation performance in terms of the Custom Score, defined earlier. Further, for each model, we considered two hyperparameters, namely, learning rate ($LR$) and momentum ($M$).
Further, viewing optimally learnt network weights, and hence the Custom (validation) Score, as functionals of the chosen hyperparameter pair, We chose the pair that maximized the said score. Test performance was reported in terms of 
the mean and the standard deviation (over the test subset across the 5 random iterations) of the corresponding classification scores measured by the chosen performance metrics.

The presence-of-eye and suitable-lighting detectors were finally arranged in the hierarchical configuration shown in Fig. \ref{fig:layer} and \ref{fig:fullflow}, tested on the hierarchically labelled dataset at hand (described in Sec. \ref{sec:data}), and the performance was reported in terms of the confusion matrix and aforementioned related indices for the final good/poor quality outcome. For any image deemed to be of poor quality, we would generate feedback on the underlying quality issue (see Fig. \ref{fig:fullflow}).

\section{Experimental Results \& Discussion} 

\begin{figure}[!t]
    \centering 
    \vspace*{-0.5cm}
\includegraphics[width=0.9\linewidth]{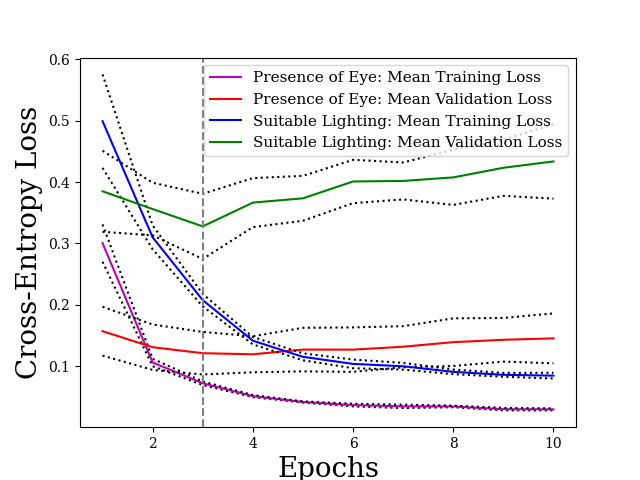}
\caption{Cross-validation loss plot for ResNet-18 (all layers trained without transform): (a) Presence-of-eye detector; (b) Suitable-lighting detector. Dotted lines bounding each loss curve denote $\textit{mean} \pm \textit{standard deviation}$.
}
\label{fig:loss}
\end{figure}


For the presence-of-eye detector, the test performance was superior when all layers were trained (compared to when only last FC layer was trained, see Table \ref{tab:pres}). 
For the suitable-lighting detector, the test performance of the variant where no transform was applied turned out to be superior in comparison to the one with Wavelet transform.
In either case, the superiority was observed across all performance indices. For completeness, we furnished optimal hyperparameters case-wise.
Further, for each detector, an optimal variant minimized the validation loss at epoch 3 (with different and somewhat significant variances) as shown in Fig. \ref{fig:loss}. 




The hierarchical classification test shows an overall strong end-to-end performance characterized by high values along the diagonal of the classification confusion matrix (table \ref{tab:hier}), with an especially strong affinity for identifying high-quality images (eye present and well-lit, ground truth $PL$) - images passing both tier 1 (eye presence) and tier 2 (satisfactory lighting). The confusion matrix shows that tier 1 (tasked with flagging all images with eye not present $\overline{P}$) works with near-perfect performance. Majority of the misclassifications occurred because eye-present and badly-lit images ($P\overline{L}$) were misclassified as eye-present and well-lit ($PL$) which can be attributed to the need for improvement in tier 2.
Grouping $\overline{P}$ and $P\overline{L}$ together as low-quality and taking $PL$ as high-quality, we can view this as a binary classification result for quality assessment. For this binary classification, we obtain the performance metric values to be 91\%, 0.9037, and 0.8244 for accuracy, P4 and Custom score respectively.

\begin{table}[t!]
    \centering
    \caption{Hierarchical classification confusion matrix for 'presence of eye' and 'lighting' features given 100 samples per ground truth label (fraction in parenthesis)}
    \begin{tabular}{c|ccc}
    \toprule 
        \textbf{Ground Truths} & \textbf{Predicted $\overline{P}$} & \textbf{Predicted $P\overline{L}$} & \textbf{Predicted $PL$} \\
    \midrule
        $\overline{P}$ & 99 $(0.99)$ & 0 $(0)$ & 1 $(0.01)$ \\
        $P\overline{L}$ & 1 $(0.01)$ & 73 $(0.73)$ & 26 $(0.26)$ \\
        $PL$ & 0 $(0)$ & 0 $(0)$ & 100 $(1)$ \\
    \bottomrule
    \end{tabular}
    \label{tab:hier}
\end{table}



While related eyecare research usually attempt AI-based detection of specific eye diseases, such as Diabetic Retinopathy \cite{retino}, development of a general quality assessment tool that apply to a broad spectrum of eye diseases has hitherto remained relatively less explored, possibly due to 
unavailability of disease-specific data for many diseases. In contrast and as a remedy, we approached the problem not from a disease-centric perspective but from an ocular structural perspective. The architecture for the above experiments was primarily adopted from works in eye-tracking \cite{eyenet}, due to its focus on the eye structure within images, and majorly influenced by other AI-based biomedical solutions.
In particular, this preliminary study was an effort toward assisting every teleconsultation with an AI-based tool that assesses captured images, possibly repeatedly and keep the subject ready with sufficiently high-quality ones, so that time and effort of the clinician is optimally utilized. The objective would be fully realized as we develop a comprehensive solution in future.

\bibliography{refs}
\bibliographystyle{ieeetr}
\end{document}